\documentstyle[12pt]{article}
\newcommand{\be}{\begin{equation}}
\newcommand{\ee}{\end{equation}}
\newcommand{\beq}{\begin{eqnarray}}
\newcommand{\eeq}{\end{eqnarray}}

\begin{document}
\def\a{\alpha}
\def\b{\beta}
\def\g{\gamma}
\def\d{\delta}
\def\e{\epsilon}
\def\l{\lambda}
\def\L{\Lambda}
\def\m{\mu}
\def\n{\nu}
\def\p{\pi}
\def\r{\rho}
\def\s{\sigma}
\def\t{\tau}
\def\f{\phi}
\def\k{\kappa}
\def\x{\xi}
\def\w{\omega}
\def\W{\Omega}
\def\ap{\alpha^\prime}
\def\intz{\int d^2z\;}
\def\c{\frac{1}{4 \pi}}
\def\de{\partial}
\def\deb{\bar{\partial}}
\def\Ab{\bar{A}}
\def\xt{\tilde{x}}
\def\yt{\tilde{y}}
\def\zt{\tilde{z}}
\def\xh{\hat{x}}
\def\xx{\vec{X}}
\def\Xt{\tilde{X}}
\def\Xh{\hat{X}}
\def\Pt{\tilde{P}}
\def\Ph{\hat{P}}
\def\et{\tilde{e}}
\def\eh{\hat{e}}
\def\Et{\tilde{E}}
\def\Eh{\hat{E}}
\def\Gt{\tilde{G}}
\def\Bt{\tilde{B}}
\def\ft{\tilde{\phi}}
\def\bra{\langle}
\def\ra{\rightarrow}
\def\ket{\rangle}
\def\diag{{\rm diag}}
\titlepage
\begin{flushright}
CERN-TH.6960/93 \\
ROM2F/93/24 \\
hep-th/9308112
\end{flushright}
\vspace{7ex}
\begin{center}
{\bf A PROBLEM WITH NON-ABELIAN DUALITY?} \\
\vspace{4ex}
M.~Gasperini${}^{(a, b)}$, R. Ricci${}^{(a, c)}$ and G.~Veneziano${}^{(a)}$ \\
\vspace{8mm}
${}^{(a)}$
{\em Theory Division, CERN, Geneva, Switzerland} \\
${}^{(b)}$
{\em Dipartimento di Fisica Teorica, Universit\`a di Torino \\
and INFN, Sezione di Torino, Turin, Italy} \\
${}^{(c)}$
{\em Dipartimento di Fisica, Universit\`a di Roma ``Tor Vergata'' \\
and INFN, Sezione di Roma 2, Rome, Italy} \\
\vspace{5ex}
{\small {\bf  ABSTRACT}}
\end{center}
{\small
We investigate duality transformations in a class of
backgrounds with  non-Abelian isometries, i.e.
Bianchi-type (homogeneous)
cosmologies in arbitrary dimensions.
Simple duality transformations for the metric and the antisymmetric
tensor field, generalizing those known
from the Abelian isometry (Bianchi I) case,
are obtained using either a Lagrangian or a Hamiltonian
approach. Applying these prescriptions to a
specific conformally invariant
$\s$-model,
we show that no dilaton transformation
leads to a new conformal background. Some possible
ways out of the problem are suggested.}
\vspace{10 mm}
\vfill
\begin{flushleft}
CERN-TH.6960/93 \\
ROM2F/93/24 \\
July 1993 \end{flushleft}
\section{Introduction}

Duality transformations on conformal string backgrounds have
recently attracted considerable attention. In a restricted sense
duality transformations connect two (or more) apparently
different, but actually equivalent,  string theories.
In the generalized sense
used in this paper, duality transformations are simply meant to connect
a given conformal background to other, generally inequivalent,
conformal backgrounds.
Examples of the latter type of dualities are the $O(d,d;R)$ Narain
\cite {NSW}
transformations connecting all possible toroidal compactifications in
$d$ dimensions. By contrast, just
an $O(d,d;Z)$ subgroup relates physically equivalent  theories \cite {GRV}.

An interesting feature of  restricted duality transformations
in the case of toroidal compactifications is the necessity to
accompany the change in the metric and torsion fields by a
suitable change of the dilaton. Only then can strict duality
hold to all orders in the string-loop expansion \cite {ALV}.

In the case of homogeneous, Bianchi I cosmological backgrounds,
a generalization of Narain's group can be defined \cite {V, T, MV, S}.
Interestingly enough,  it relates cosmologies of the
standard kind (FRW   Universe undergoing a decelerating expansion)
to inflationary cosmologies offering some hope
that, in string theory, inflation can be incorporated naturally
in a pre-Big-Bang phase \cite {PBB}. In this case, even the weaker form of
duality requires, for the maintenance of conformal invariance,
a non-trivial dilaton transformation. In fact, it is
the very presence of a time-dependent dilaton that makes it possible
for inflationary solutions to exist.

The presence of an $O(d,d;R)$ group is not at all confined to
cosmological backgrounds. It was indeed shown \cite {S} that it is a general
property   of backgrounds  possessing $d$ Abelian
isometries. In the case of Abelian isometries it is also possible
to understand \cite {T, Dil} why and how the dilaton is to be transformed in
order that the quantum equivalence of duality-related theories holds
or, more generally, that the original conformal invariance is not lost.

Generalizing the above construction to backgrounds with
non-Abelian isometries is an obvious mathematical challenge \cite{FTFJ}.
At the
same time, such a problem is of great physical interest in a
cosmological context since some of the most interesting cosmological
models have non-Abelian rather than Abelian isometries.
One of the fundamental problems cured by inflation, the flatness
problem, cannot be even addressed without considering
cosmologies with non-Abelian isometries (Bianchi IX and V).

Some time ago,  in a very stimulating
paper \cite{Que}, de la Ossa and Quevedo (DOQ) proposed  a possible way
to implement duality
transformations for backgrounds with
non-Abelian  isometries  (``non-Abelian duality'', for short).
Besides giving
a method for computing the transformation of the metric and
antisymmetric tensors, these authors also gave a recipe for
determining the dilaton transformation, checking its validity through
$\beta$-function calculations in some cases.

In this paper we shall apply the DOQ method to the case of general
Bianchi-type models (most general homogeneous cosmologies in arbitrary
dimensions) double-checking their transformations of $G$ and $B$
by a completely different (Hamiltonian) approach.
However, and to our surprise, when we add to these the dilaton transformation
as given by DOQ, we fail to satisfy, in a specific example, the
$\beta$-function constraints. What is worse, we find that no other
dilaton transformation is capable of restoring the vanishing
of all $\beta$-functions. Some possible interpretations of this
surprising result are given at the end.

\section{Strings in homogeneous cosmological backgrounds}
The two-dimensional $\s$-model action describing string
propagation in a generic background of its massless modes
can be written (in orthonormal gauge) as:
\be
S = \c \intz \{\de X^M [G_{MN}(X) + B_{MN}(X)] \deb X^N + \frac{\sqrt{g}}{2}
R^{(2)}\f(X)\},
\qquad M, N = 0, \ldots, d,
\label{1}
\ee
where $X^M \equiv (t, X^m)$ ($m = 1, \ldots, d$) are the string coordinates,
$R^{(2)}$ is the scalar curvature of the $2$-dimensional world-sheet
and
the fields $G_{MN} = G_{NM}$, $B_{MN} = -B_{NM}$ and $\f$ are general functions
 of
$X$.
Having in mind possible cosmological applications,
we shall restrict ourselves
to the particular case of spatially homogeneous cosmological
backgrounds, for which a synchronous frame can be defined
in which
\be
G_{MN}(X) = \pmatrix{-1 & 0 \cr 0 & G_{mn}(t, \xx) \cr}, \qquad
B_{MN}(X) = \pmatrix{ 0 & 0 \cr 0 & B_{mn}(t, \xx) \cr}.
\label{2}
\ee
The requirement of
spatial homogeneity implies that the $d$-dimensional spatial submanifold
is invariant under the action of a $d$-parameter transitive
Lie group of isometries $\cal{G}_d$.
The generators $T_\a$ of the corresponding Lie algebra
are expressible in terms of a set of  $d$-dimensional
Killing vectors $\x_\a^m$,
which can be taken to depend only on $\xx$:
\be
T_\a = \x_\a^m(\xx) \de_m, \qquad \a = 1, \ldots, d.
\label{3}
\ee
Lie algebras corresponding to different groups
are fully characterized by their structure constants
$C_{\a\b}^\g$, defined as usual  by
\be
[T_\a, T_\b] = C_{\a\b}^\g T_\g.
\label{3bis}
\ee
A complete classification of the allowed algebras exists
for the phenomenologically interesting case of
$d = 3$ (see, for instance, \cite{Bianchi}).
Accordingly, all four-dimensional spatially
homogeneous spacetimes, also known as Bianchi models,
fall into one of nine classes. Bianchi I, characterized by
an Abelian isometry group isomorphic to the three-dimensional
translation group
(all $C_{\a\b}^\g = 0$), coincides with the spatially flat anisotropic
Friedmann-Robertson-Walker (FRW) Universe, while
Bianchi V (respectively IX)   contains, as a special
case, the FRW open (respectively closed) isotropic Universe.

The existence of isometries allows one to factorize the
  background fields $G$ and $B$ in the form:
\be
G_{mn}(t, \xx) = e_m^\a(\xx)
\g_{\a\b}(t) e_n^\b(\xx), \qquad \g_{\a\b} = \g_{\b\a},
\label{4}
\ee
\be
B_{mn}(t, \xx) = e_m^\a(\xx) \b_{\a\b}(t)
e_n^\b(\xx), \qquad \b_{\a\b} = -\b_{\b\a},
\label{5}
\ee
where all dependence on the spatial coordinates is contained in the
``triads'' $e_m^\a$.  The specific form of the latter is fixed,
up to space diffeomorphisms, by the particular
isometry group involved. The   linear
differential operators
\be
  e^m_\a \de_m \; \; , \;\; e_m^\a e^n_\a \equiv \d_m^n,
\label{5b}
\ee
satisfy the same commutation relations as the generators $T_\a$.
Thus, eq.~(\ref{3bis}) can be equivalently rewritten as \cite{Landau}
\be
e_\a^m e_\b^n ( \de_n e_m^\g - \de_m e_n^\g) = C_{\a\b}^\g.
\label{5a}
\ee
Combining (\ref{2}),(\ref{4}) and (\ref{5}) into (\ref{1}) we
obtain
\be
S = \c \intz \{-\de t \deb t + \de X^m e_{m}^{\a}(\g_{\a\b} + \b_{\a\b})
e_{n}^{\b} \deb X^n\ + \frac{\sqrt{g}}{2}R^{(2)}\f\},
\label{6}
\ee
which will be our starting point for discussing non-Abelian duality.
\section{Non-Abelian duality}
Starting from (\ref{6}), a ``dual''
$\s$-model action can be defined with respect to the full
non-Abelian isometry group $\cal{G}_d$, in strict analogy
with the Abelian case.
We shall follow, successively, a Lagrangian and a Hamiltonian approach,
showing that both yield identical results.

\subsection{Lagrangian approach}
In the Lagrangian approach \cite{Dil, Kir, Que}
duality transformations for the background tensor fields
are obtained from a chain of formal manipulations on the
functional integral which defines the partition function of the
initial theory:
\be
\cal{Z} = \int [dt][dX^m]e^{- S[t, X^m]}.
\label{7}
\ee
As a first step, one gauges the global symmetry corresponding to $\cal{G}_d$
by introducing a set of pure-gauge potentials $A^\g$, $\Ab^\g$,
which are minimally coupled to the string coordinates:
\be
\de X^m \ra \de X^m + A^\g \x_\g^m(\xx), \qquad
\deb X^m \ra \deb X^m + \Ab^\g \x_\g^m(\xx).
\label{8}
\ee
The new total action $S'$ reads:
\beq
S' & = & S +
\c \intz \{
A^\g \x_\g^m e_{m}^{\a}(\g_{\a\b} + \b_{\a\b}) e_{n}^{\b} \deb X^n +
\nonumber \\
& & \Ab^\d \de X^m e_{m}^{\a}(\g_{\a\b} + \b_{\a\b}) e_{n}^{\b} \x_\d^n +
\nonumber \\
& & A^\g \Ab^\d \x_\g^m e_{m}^{\a}(\g_{\a\b} + \b_{\a\b}) e_{n}^{\b} \x_\d^n +
\Xt_\g F^\g \},
\label{9}
\eeq
where $F^\g$ is the field-strength
corresponding to $A^\g$ and $\Ab^\g$ and the Lagrange multipliers
$\Xt_\g$ are used to
enforce the constraint $F^\g =0$.
In terms of the action $S'$ we have:
\be
\cal{Z} = \int [dt][dX^m][dA^\a][d\Ab^\b][d\Xt_\g] \frac{1}{V_{\cal{G}_d}}
e^{-S'[t, X^m, A^\a, \Ab^\b, \Xt_\g]},
\label{10}
\ee
where $V_{\cal{G}_d}$ stands for the formal gauge group volume, it being
understood that a Faddeev-Popov gauge-fixing procedure must be performed to
render the path-integral well-defined.
The original model is recovered from (\ref{10}) by first integrating over the
Lagrange multipliers and then fixing the potentials to zero.
Alternatively, the fact that the action is quadratic in the gauge
potentials allows one to obtain the dual theory by integrating first over $A$
 and
$\Ab$ and by subsequently fixing the residual gauge symmetry in a suitable way.
It is convenient to rewrite $S'$ in the compact form:
\be
S' = S + \c \intz (A^\g \bar{u}_\g + \Ab^\d u_\d + A^\g m_{\g\d} \Ab^\d),
\label{11}
\ee
where
\beq
u_\d & = & - \de \Xt_\d +
        \de X^m e_{m}^{\a}(\g_{\a\b} + \b_{\a\b}) e_{n}^{\b} \x_\d^n \\
\bar{u}_\g & = & \deb \Xt_\g +
        \x_\g^m e_{m}^{\a}(\g_{\a\b} + \b_{\a\b}) e_{n}^{\b} \deb X^n, \\
m_{\g\d} & = & C_{\g\d}^\l \Xt_\l
        + \x_\g^m e_{m}^{\a}(\g_{\a\b} + \b_{\a\b}) e_{n}^{\b} \x_\d^n.
\label{12}
\eeq
After (classical) integration over the gauge potentials one gets:
\be
S'' = S - \c \intz [u_\g (m^{-1})^{\g\d} \bar{u}_\d].
\label{13}
\ee
A convenient gauge choice, whose viability is locally guaranteed
by the transitiveness of the group
$\cal{G}_d$, turns out to be $X^m \equiv C^m$, with $C^m$ a suitable
(possibly group-dependent) constant vector.
Under this choice (\ref{13})
simplifies considerably, owing to the property
$\left. e_{n}^{\b} \x_\d^n \right|_{\xx =
\vec{C}} = \d_\d^\b$.
This yields the following general form for the  dual action:
\be
\tilde{S} = \c \intz \{-\de t \deb t + \de \Xt_\g [(\g +
\b + \k)^{-1}]^{\g\d} \deb \Xt_\d + \frac{\sqrt{g}}{2}R^{(2)}\f\},
\label{14}
\ee
where $\k$ stands for the antisymmetric matrix defined by
\be
\k_{\a\b} \equiv C_{\a\b}^\g \Xt_\g.
\label{14b}
\ee
{}From eq.~(\ref{14}), the following prescription for the dual backgrounds
$\Gt$ and $\Bt$ can be inferred:
\be
\Gt + \Bt = (\g + \b + \k)^{-1},
\label{15}
\ee
or, using the symmetry properties of $\g$, $\b$ and $\k$,
\beq
\Gt & = & (\g - \b - \k)^{-1} \g (\g + \b + \k)^{-1},
\label{15a} \\
\Bt & = & -(\g - \b - \k)^{-1}(\b + \k)(\g + \b + \k)^{-1}.
\label{15b}
\eeq
We note that the above transformations correctly reduce to the
Abelian ones when $\k = 0$.
\subsection{Hamiltonian approach}
The transformation rules (\ref{15a}) and (\ref{15b})
for the background tensor fields can be
alternatively inferred from a Hamiltonian framework.
The total Hamiltonian density for a string in the backgrounds (\ref{2})
can be written as:
\be
H_T = \frac{1}{2} {X'}^0 G_{00} {X'}^0 + \frac{1}{2}
P_0 G^{00} P_0 + H, \qquad H = \frac{1}{2} Z^I M_{IJ} Z^J,
\label{20}
\ee
where
\be
Z^I \equiv (P_i, X'^i), \qquad I = 1, \ldots, 2d,
\qquad i = 1, \ldots, d
\label{21}
\ee
are $2d$-dimensional phase-space coordinates
and $M$ is the $2d \times 2d$ matrix \cite{GRV}
\be
M = \pmatrix{ G^{-1} & -G^{-1}B \cr BG^{-1} & G -BG^{-1}B \cr}.
\label{22}
\ee
We shall try to follow the  strategy of \cite {GRV}  in the
case  of
$X^i$-dependent, spatially homogeneous
backgrounds (\ref{4}), (\ref{5}).
In this case $M$ reads:
\be
M_{IJ} = \pmatrix{ e^i_\a \g^{\a\b} e_\b^j &  - e^i_\a \g^{\a\l} \b_{\l\b}
e^\b_j \cr
e_i^\a \b_{\a\l} \g^{\l\b} e_\b^j & e_i^\a (\g_{\a\b} - \b_{\a\l} \g^{\l\m}
\b_{\m\b}) e^\b_j \cr}.
\label{24}
\ee
We now perform   two successive
classical canonical transformations.
The first one is induced by the   generating
functional:
\be
F = - \int d\s \: d\t \; (X'^m e^\a_m \d_{\a\b} \eh_n^\b \Xh^n),
\label{25}
\ee
where $\eh_m^\a$ is the same set of functions of the new coordinates $\Xh^i$ as
$e_m^\a$ is of $X^i$.

Introducing the new variables $E_m^\a \equiv \de_m (e_n^\a X^n)$ and
$\hat{E}_m^\a \equiv \hat{\de}_m (\eh_n^\a \Xh^n)$ we can write, after
use of (\ref{5a}) and
some algebra,
\beq
X'^n e_{n\a} & = & \Ph_m \Eh^m_\a
\label{27a} \\
P_n  e_{\a}^n & = & \Eh_{m\a} \Xh'^m - C_{\a\b}^\g \Ph_m \Eh^{m\b} \eh_{\g r}
\Xh^r.
\label{27b}
\eeq
The second canonical transformation is just a general coordinate
transformation:
\be
\Xt_\a = \eh_{\a n} \Xh^n, \qquad \Pt_\a = \Ph_n \Eh_\a^n.
\label{28}
\ee
Combining the two
transformations, we can luckily re-express the quantities
appearing in the Hamiltonian entirely in terms of
the final phase-space coordinates $\Xt'$ and $\Pt$:
\beq
X'^n e_{n\a} & = & \Pt_\a
\label{29a} \\
P_n  e_{\a}^n & = & \Xt'_\a - C_{\a\b}^\g \Pt_\b \Xt_{\g}.
\label{29b}
\eeq
Substituting eqs.~(\ref{29a}) and
(\ref{29b}) into the
expression for $H$, we can reinterpret the new Hamiltonian as
one defining a new background matrix $\tilde{M}$ given by:
\be
\tilde{M} = \pmatrix{ (\g + \b + \k) \g^{-1} (\g - \b - \k) &
(\b + \k) \g^{-1}\cr
\g^{-1} (\b + \k) &
\g^{-1} \cr}.
\label{30}
\ee
After use of  eq.~(\ref{22}),  one finds that (\ref{30}) defines
the same transformations on
$\Gt$ and $\Bt$ as the one obtained in the Lagrangian approach
(eqs.~(\ref{15a}) and (\ref{15b})).

We close this section by noticing that the dual backgrounds
(\ref{15a}) and (\ref{15b}) do not share in general the same isometries as the
original ones. This result is in agreement with the general conclusions of
\cite{Que}.
\section{Transformation of the dilaton and a puzzle}
As we discussed already in the introduction, the maintenance of
conformal invariance requires, even in the Abelian case,
a non-trivial
transformation of the dilaton. The necessity of such a transformation
is easily inferred from the computation of  $\b$-functions in
the original and duality-related $\s$-models.  In the Lagrangian approach,
a more direct
method for determining how the dilaton has to transform runs as
follows \cite {T, Dil}.

In the formal functional-integral manipulations used to go from
one theory to its dual we have been cavalier about functional
determinants, in particular about those coming from
integration over the gauge potentials.
Indeed,
since the coefficient of the term quadratic in the potentials is
not an elliptic differential operator,
the corresponding functional integral is ill-defined and needs to be
regulated.

The functional determinant can then be explicitly calculated using heat
kernel techniques \cite{T, Dil}.
In particular, its Weyl anomaly part provides an
additional term proportional to $R^{(2)}$ to the dual action. This is
naturally interpreted as a shift of the original dilaton field.

In the case of Abelian isometries one finds:
\be
\tilde{\f} = \f - \ln \det(G).
\label{40}
\ee

Eq.~(\ref{40}) can be also obtained in the context of the
dimensionally-reduced
string effective action \cite{V, T, MV}.
The same quantum contribution to the dilaton transformation
is also expected to follow in the
Hamiltonian approach, provided one correctly takes
into account the problems of
operator ordering when writing down the quantum version of  canonical
transformations. This has not yet been done explicitly,
to our knowledge.

The following question  naturally arises at this point:
is there a transformation of the dilaton
which guarantees  conformal invariance
in the non-Abelian case? And if yes, which is it?

It has been argued  \cite{Que} that the correct
prescription is simply:

\be
\ft = \left[\f + \frac{1}{2} \ln \frac{\det\Gt}{\det G} - \ln\det
(\frac{\d \cal{F}}{\d \w})\right]_{\cal{F} = 0}.
\label{41}
\ee
In eq.~(\ref{41}) $\cal{F}$ is the gauge
fixing function appearing in the path-integral representation of the partition
function and $\w$ are the parameters of the isometry transformation.
Eq.~(\ref{41}) was shown \cite{Que} to correctly reinstate one-loop conformal
invariance
in the case of $\s$-models with maximal $SO(d)$ isometry symmetry.
This is a somewhat special case, however, since no $B$
field is introduced by the duality transformation if it was initially zero.

Since no rigorous argument exists ensuring the general
validity of eq.~(\ref{41}),
we made an explicit check of that recipe
in the case of a
particular, Bianchi-type conformal background.
Consider the $\s$-model defined by (\ref{6}) with $d = 3$,
$B_{mn} = \f \equiv 0$, and

\be
G_{mn}(t, x, y, z) = \diag (a^2(t), a^2(t) e^{-2x}, a^2(t) e^{-2x}).
\label{43}
\ee
The metric (\ref{43}) is of the form (\ref{4}) with
\be
e_m^\a = \diag(1, e^{-x}, e^{-x}), \qquad
\g_{\a\b} = a^2(t) \d_{\a\b}.
\label{44}
\ee
Using (\ref{5a}) one obtains,
for the non-vanishing structure constants:
\be
C_{12}^2 = -C_{21}^2 = C_{13}^3 = -C_{31}^3 = 1, \qquad \mbox{zero otherwise}.
\label{45}
\ee
showing that the model is of the  Bianchi V type.

It is easy to see that, by choosing $a(t) \equiv t$,
the above background fulfils the $\b$-function
equations trivially (after adding the right number of ``spectator''
dimensions).
Indeed  the whole Riemann
tensor of the model vanishes identically, implying that $G_{MN}$ is flat
(it provides, indeed,
an unconventional parametrization of the ``Milne'' \cite{Milne} portion of
Minkowski space-time). For an analogous abelian case see \cite{GMV}.

The new tensor backgrounds $\Gt$ and $\Bt$, when
calculated by means of
eqs.~(\ref{15a}) and (\ref{15b}), read (hereafter we drop the tilde for
the new spatial coordinates):
\be
\Gt = \frac{1}{D} \pmatrix{ t^4 & 0 & 0 \cr
                0 & t^4 + z^2 & -yz \cr
                0 & -yz & t^4 + y^2 \cr}, \qquad
\Bt = \frac{1}{D} \pmatrix{ 0 & -t^2y & -t^2z \cr
                t^2y & 0 & 0 \cr
                t^2z & 0 & 0 \cr},
\label{47}
\ee
where $D = t^2(t^4 + y^2 + z^2)$.
The prescription (\ref{41}), adapted to our case,
gives:
\be
\ft = -\ln(\g + \k) = -\ln(D).
\label{48}
\ee
By direct computation one can see that
eqs.~(\ref{47}) and (\ref{48}) do not satisfy the $\b$-function equations,
in particular the one for the
$B$ field ($H_{MAB} = \de_{[M}B_{AB]}$)
\be
\de_M\left(e^{-\ft}\sqrt{\det\Gt} \tilde{H}^{MAB}\right) = 0.
\label{50}
\ee
One finds, indeed, for the transformed backgrounds,
\be
\tilde{H}^{201} = -\frac{4y}{t}, \qquad \tilde{H}^{301} = -\frac{4z}{t},
\qquad e^{-\ft}\sqrt{\det\Gt} = t^3,
\label{51}
\ee
so that the $A=0$, $B=1$ component of eq.~(\ref{50}) is clearly not satisfied.

In order to see whether conformal invariance can be recovered  by just
changing the definition of the transformed dilaton, we treated $\ft$ as an
unknown function in eq.~(\ref{50}). The general solution
for $\ft$ turns out to be:
\be
\ft = -\ln\frac{t^4 + y^2 + z^2}{yz} + f\left(\frac{y}{z}\right),
\label{52}
\ee
with $f$  an arbitrary function of $y/z$.
The dilaton $\b$-function equation requires $f$ to satisfy a Riccati-type
differential equation.
Unfortunately, independently of the choice of $f$, eq.~(\ref{52}) does not
fulfil the remaining $\b$-function equations
\be
\tilde{R}_A^B +\nabla_A \nabla^B\ft -
\frac{1}{4}\tilde{H}_{AMN}\tilde{H}^{BMN} = 0.
\label{55}
\ee
Consider for instance the $A=0$, $B=0$ component, which does not depend on
$f$. Using the expressions (\ref{47}) and (\ref{52}) for the background fields
one finds
\be
\tilde{R}_0^0 = 2\frac{(y^2 + z^2)^2  - 8t^4 (z^2 + y^2) + 3t^8}{t^2(z^2
+ y^2 + t^4)^2},
\label{56}
\ee
\be
\nabla_0 \nabla^0 \ft = -t^2\frac{4t^4 - 12(y^2 + z^2)}{(z^2 + y^2 + t^4)^2},
\label{57}
\ee
\be
-\frac{1}{4}\tilde{H}_{0AB}\tilde{H}^{0AB} = 8t^2\frac{y^2 + z^2}{(z^2
+ y^2 + t^4)^2},
\label{58}
\ee
which add up to $2/t^2 \ne 0$. Thus no choice for the transformed dilaton
appears to restore conformal invariance.

Before discussing the possible implications of this result, we have to dismiss
the possibility that the violation of conformal invariance we found can be
fixed by higher order terms in the $\b$-functions. The following argument shows
that this is impossible:
let us change the time coordinate from cosmic time $t$ to $\tau = \sqrt{t}$. It
is easy to check that the transformed backgrounds become homogeneous of degree
$-1$ in the new coordinates. As a result, one can show that the contributions
to the $\b$-functions of each extra $\s$-model loop contain an extra power of
$X^{-1}$ (where $X$ stands for any one of the new coordinates $y$, $z$ or
$\tau$). Since the violation of conformal invariance that we found is of
$O(X^{-1})$, it cannot be cancelled by higher order terms.
\section{Discussion}

Barring some trivial computational mistake, what could be the meaning of
our counterexample? Obviously, we have no definite answer to this question.
All we can offer at the moment are some conjectures which we list now in order
of decreasing (subjective) appeal.
\begin{enumerate}
\item In the non-Abelian case, the functional determinant encountered
in going from one theory to its dual appears to be more
 complicated than in the
Abelian case. As already observed by
Schwarz and Tseytlin \cite{Dil},
counterterms of types other than the dilaton's can be induced a priori.
In the Abelian case these happen to vanish. Unfortunately,
even in our case, they do not seem to be of much help since,
as shown by a simple dimensional counting, they would be of
higher order in $\ap$. By contrast the non-Abelian correction
to the duality transformations, in spite of involving derivatives
of the backgrounds through $C^\a_{\b\gamma}$, are easily shown
not to contain any extra factors of $\ap$.
\item Entirely new counterterms could be generated, i.e.
operators that do not correspond to any of the massless backgrounds.
Possible examples could be a tachyonic background or an excited massive
background. If this is the case, then presumably more
 and more massive fields
will be brought in by the duality transformation as one goes to higher
and higher orders in $\a ^{'}$.
\item The determinant could generate non-local counterterms, in
which case the dual theory would have no standard
 ``string-in-a-background''
interpretation but only a conformal field theory formulation.
\item Finally, there could simply be no conformal field theory
which is dual to one with a particularly complicated non-abelian
isometry.
\end{enumerate}
In conclusion, and independently of what the solution of our puzzle will
eventually be, we do feel
that a full understanding of non-Abelian duality
represents a very interesting physical and mathematical challenge.

\vskip .5cm

We are grateful to J.~Maharana for an independent check of some of our
calculations and for discussions. We also acknowledge useful conversations
with L.~Alvarez Gaum\'e, X.~de la Ossa, A.~Giveon, E.~Kiritsis,
 F.~Quevedo,
E.~Rabinovici, A.~Sagnotti and A.~A.~Tseytlin.
One of us (R.~R.) wishes to thank the Dipartimento
 di Fisica, Universit\`a di
Roma ``Tor Vergata'' for partial financial support.


\begin{thebibliography}{99}

\bibitem{NSW} K.~S.~Narain, {\em Phys. Lett.\/} {\bf 169B} (1986) 41; \\
K.~S.~Narain, M.~H.~Sarmadi and E.~Witten,
{\em Nucl. Phys.\/} {\bf B279} (1987) 369.
%
\bibitem{GRV} A.~Giveon, E.~Rabinovici and G.~Veneziano, {\em Nucl. Phys.\/}
{\bf B322} (1989) 167; \\
A.~Shapere and F.~Wilczek, {\em Nucl. Phys.\/} {\bf B320} (1989) 669.
%
\bibitem{ALV} P.~Ginsparg and C.~Vafa, {\em Nucl. Phys.\/} {\bf B288}
(1987) 414; \\
E.~Alvarez and M.~Osorio, {\em Phys. Rev.\/} {\bf D40} (1989)
1150.
%
\bibitem{V} G.~Veneziano, {\em Phys. Lett.\/} {\bf B265} (1991) 287.
%
\bibitem{T} A.~A.~Tseytlin, {\em Mod. Phys. Lett.\/} {\bf A6} (1991) 1721.
%
\bibitem{MV} K.~A.~Meissner and G.~Veneziano, {\em Phys. Lett.\/}
{\bf B267} (1991) 33;
{\em Mod. Phys. Lett.\/} {\bf A6} (1991) 3397; \\
M.~Gasperini and G.~Veneziano, {\em Phys. Lett.\/} {\bf B277} (1992) 256.
%
\bibitem{S} A.~Sen, {\em Phys. Lett.\/} {\bf B271} (1991) 295;
ibid. {\bf B274} (1992) 34; \\
S.~F.~Hassan and A.~Sen, {\em Nucl. Phys.\/} {\bf B375} (1992) 103; \\
J.~Maharana and J.~H.~Schwarz,
 {\em Non Compact Symmetries in String Theory\/},
Caltech preprint, CALT-68-1790 (1992).
%
\bibitem{PBB} M.~Gasperini and G.~Veneziano,
{\em Astroparticle Phys.\/} {\bf 1} (1993) 317.
%
\bibitem{Dil} T.~H.~Buscher {\em Phys. Lett.\/} {\bf B201} (1988) 466; \\
E. Kiritsis, {\em Mod. Phys. Lett.\/} {\bf A6} (1991) 2871;\\
M.~Ro\v{c}ek and E.~Verlinde,
 {\em Nucl. Phys.\/} {\bf B373} (1992) 630; \\
A.~Giveon and M.~Ro\v{c}ek, {\em Nucl. Phys.\/} {\bf B380} (1992) 128; \\
A.~S.~Schwarz and A.~A.~Tseytlin, {\em Nucl. Phys.\/},
 {\bf 399} (1993) 691.
%
\bibitem{FTFJ} B.~E.~Fridling and A.~Jevicki, {\em Phys. Lett.\/} {\bf 134B}
(1984) 70; \\
E.~S.~Fradkin and A.~A.~Tseytlin, {\em Ann. Phys.\/} {\bf 162} (1985) 31.
%
\bibitem{Que} X.~C.~de la Ossa and F.~Quevedo,
{\em Nucl. Phys.\/} {\bf B403} (1993) 377.
%
\bibitem{Bianchi} A.~H.~Taub, {\em
Ann. Math.\/} {\bf 53} (1951) 472; \\
M.~P.~Ryan Jr. and L.~C.~Shepley, {\em Homogeneous Relativistic Cosmologies\/},
(Princeton University Press, Princeton 1975).
%
\bibitem{Landau} L.~D.~Landau and E.~M.~Lifshits, {\em The Classical Theory of
Fields\/}, (Pergamon Press, 1987).
%
\bibitem{Kir} E.~Kiritsis, {\em Exact duality symmetry in CFT and String
Theory\/}, CERN Preprint, CERN-TH 6797/93.
%
\bibitem{Milne} E.~A.~Milne, {\em Nature\/} {\bf 130} (1932) 4.
%
\bibitem{GMV} M.~Gasperini, J.~Maharana and G.~Veneziano,
 {\em Phys. Lett.\/}
{\bf B272} (1991) 277.
%
\end{thebibliography}
\end{document}